\documentclass[prl,aps,showpacs,nofootinbib,superscriptaddress,tightenlines,twocolumn]{revtex4}

\usepackage{epsfig}
\usepackage{amssymb,latexsym,amsmath}
\newcommand{\eq}[1]{\begin{align} #1 \end{align}}

\begin{document}

\title{Statistical Ensembles with Fluctuating Extensive Quantities}

\author{M.I. Gorenstein}
\affiliation{ Bogolyubov Institute for Theoretical Physics\\
  14-b, Metrolohichna str., Kiev, 03680, Ukraine}
\affiliation{Frankfurt Institute for Advanced Studies \\
  Johann Wolfgang Goethe - Universit\"at Frankfurt \\
  Ruth-Moufang-Str. 1, D-60438 Frankfurt am Main, Germany }

\author{M. Hauer}
\affiliation{Helmholtz Research School \\
  Johann Wolfgang Goethe - Universit\"at Frankfurt \\
  Ruth-Moufang-Str. 1, D-60438 Frankfurt am Main, Germany }

\begin{abstract}
We suggest an extension of the standard concept of statistical
ensembles. Namely, we introduce a class of ensembles with extensive quantities
fluctuating according to an externally given distribution. As an example the
influence of energy fluctuations on multiplicity fluctuations in limited
segments of momentum space for a classical ultra-relativistic gas is
considered.
\end{abstract}



\maketitle

Successful application of the statistical model to hadron
production in high energy collisions (see, e.g., recent
papers~\cite{stat1} and references therein) has stimulated
investigations of properties of statistical ensembles of
relativistic hadronic gases. Whenever possible, one prefers to use
the grand canonical ensemble (GCE) due to its mathematical
convenience. The canonical ensemble (CE)~\cite{CE} should be
applied when the number of carriers of a conserved charges is
small (of the order of 1), such as strange hadrons~\cite{strange},
antibaryons~\cite{antibaryons}, or charmed hadrons \cite{charm},
in otherwise large systems. The micro-canonical ensemble
(MCE)~\cite{MCE} has been used to describe small systems (with
total number of produced particles less than or equal to~10) with
additionally fixed energy, and momentum, e.g. elementary particle
collisions or annihilation. In these cases, calculations performed
in different statistical ensembles yield different results. Hence,
ensembles are not equivalent, and systems are `far away' from the
thermodynamic limit (TL).

Measurement of hadron multiplicity distributions $P(N)$ in
relativistic nucleus-nucleus collisions opens  another interesting
field of investigations. The average number of produced hadrons
ranges from $10^2$ to $10^4$, and mean multiplicities (of light
hadrons) obtained within GCE, CE, and MCE approach each other. One
refers here to the thermodynamical equivalence of statistical
ensembles and uses the GCE for fitting experimentally measured
mean multiplicities. However, the number of particles fluctuates
event-by-event. These fluctuations are usually quantified by the
ratio of variance to mean value of a multiplicity distribution
$P(N)$, the scaled variance, and are a subject of current
experimental activities. In statistical models there is a qualitative
difference in the properties of mean multiplicity and scaled variance of
multiplicity distributions. It was recently found
\cite{fluc,mce,res} that even in the TL corresponding results for
the scaled variance are different in different ensembles. Hence
the equivalence of ensembles holds for mean values in the TL, but
does not extend to fluctuations.


\vspace{0.2cm} Statistical mechanics is usually formulated through
the following steps. Firstly, one fixes the system's extensive
quantities:  volume $V$, energy $E$, momentum $\vec P$, and
conserved charges $\{Q_i\}$. Secondly, one postulates that all
microstates have equal probability of being realized. This defines
the MCE. The CE introduces temperature $T$. Each set of
microstates with fixed energy $E$ (a macrostate) is weighted with
the Boltzmann factor $e^{-E/T}$. The probability to find a
macrostate with energy $E$ in the CE is then proportional to the
number of all microstates with energy $E$ times the Boltzmann
factor $e^{-E/T}$. To define the GCE, one makes a similar
construction for conserved charges $\{Q_i\}$ and introduces
chemical potentials $\{\mu_i\}$. Canonical or grand  canonical
observables can then be obtained by averaging over the CE energy
distribution or the GCE (joint-) distribution of both energy and conserved
charges.

Fluctuations in statistical systems, e.g., multiplicity distributions
$P(N)$ in relativistic gases \cite{fluc,mce,res}, are sensitive to
conservation laws obeyed by  the system, and therefore to fluctuations of
extensive quantities. For calculation of multiplicity distributions, the choice
of statistical ensemble is then not a matter of convenience, but a physical
question. Fluctuations of extensive quantities $\vec{A}\equiv (V,E,\vec
P,\{Q_i\})$ around their average values depend not on the system's physical
properties, but rather on external conditions. One can imagine a huge variety
of these conditions, thus, MCE, CE, GCE, or pressure ensembles \cite{RR} are
only some special examples.

A more general statistical ensemble can be defined by an
externally given distribution of extensive quantities,
$P_{\alpha}(\vec{A})$. All microstates with a fixed set $\vec{A}$
are taken to be equiprobable. Thus, the probability
$P_{mce}(N;\vec{A})$ of finding the system in a macrostate with
fixed $\vec{A}$ and additionally fixed multiplicity $N$ is given
by the ratio of the number of microstates with fixed $N$ and
$\vec{A}$ to the number of microstates with fixed $\vec{A}$. The
construction of multiplicity distributions in such an ensemble
proceeds in two steps. Firstly, the MCE multiplicity distribution,
$P_{mce}(N;\vec{A})$, at fixed values of the extensive quantities
$\vec{A}$ is calculated. Secondly, this result is averaged over
the external distribution $P_{\alpha}(\vec{A})$,
\begin{eqnarray}\label{P}
P_{\alpha} (N) ~=~ \int  d\vec{A}  ~P_{\alpha}(\vec{A}) ~
P_{mce}(N;\vec{A})~.
\end{eqnarray}
The ensemble defined by Eq.~(\ref{P}), the $\alpha$-ensemble,
includes the standard statistical ensembles as particular cases.


\vspace{0.2cm} Let us illustrate above statements for a simple system of
non-interacting massless particles, neglecting the effects of quantum
statistics (Boltzmann approximation). For a particular realization of
Eq.~(\ref{P}) we choose:
\begin{eqnarray}\label{PalphaN}
P_{\alpha} (N) = \int dE ~P_{\alpha}(E) ~  P_{mce}(N;E)
\end{eqnarray}
to calculate the multiplicity distribution $P_{\alpha}(N)$ in the
presence of an energy distribution $P_{\alpha}(E)$.
We will firstly discuss the MCE multiplicity distribution
$P_{mce}(N;E)$ and then solve the integral (\ref{PalphaN}) for a
particular choice of $P_{\alpha}(E)$ in the large volume limit.

The MCE multiplicity distribution is given by~\cite{mce},
\eq{\label{Pmce} P_{mce}(N;E) ~=~\frac{1}{Z_{mce}(V,E)}~
\frac{E^{-1}~x^N}{(3N-1)!~N!}~,}
where $x\equiv gVE^3/\pi^2$, and $g$ is the particle's degeneracy
factor. The MCE partition function $Z_{mce}(V,E)$ is defined by
the normalization condition,
$\sum_{N=1}^{\infty}P_{mce}(N;E)=1$. In the TL, $x\gg 1$ in
Eq.(\ref{Pmce}),  one finds from the raw moments $\langle N^k
\rangle~\equiv~\sum_{N=1}^{\infty}~N^k~P_{mce}(N;E)$~:
\eq{\label{Nmce} \langle N\rangle  \cong
\left(\frac{x}{27}\right)^{1/4}, ~~ \omega_{mce} \equiv
\frac{\langle N^2\rangle -\langle N\rangle^2}{\langle N\rangle}
\cong \frac{1}{4}~.}
Additionally it follows from the equivalence of statistical
ensemble that GCE and MCE values for average multiplicity are
equal to each other, $\langle N \rangle_{mce} \cong \overline{N}=
gVT^3/\pi^2$, where the temperature $T$ of the GCE is found from
$E=\overline{ E}=3gVT^4/\pi^2$. Momentum spectra in the MCE also
converge to GCE Boltzmann spectra under this limit. Multiplicity
fluctuations, expressed by the scaled variances, are however
different, $\omega_{gce}= 1$ and $\omega_{mce}=1/4$.

In the TL the MCE distribution (\ref{Pmce}) converges to a Gaussian,
\eq{\label{PmceG} P_{mce} (N;E) \cong \frac{1}{\sqrt{2\pi
\omega_{mce}\langle
  N\rangle}} \exp\left[- \frac{(N-\langle
    N\rangle)^2}{2\omega_{mce}\langle N\rangle}\right].
}
The Normal form of distributions is a general
feature of all statistical ensembles in the TL \cite{clt}.


\vspace{0.2cm}
The GCE energy distribution is equal to:
\begin{equation}\label{PgceE}
P_{gce}(E) = \frac{1}{Z_{gce}(V,T)}~ \exp\left(-\frac{E}{T}\right)
Z_{mce}(V,E)~,
\end{equation}
where the GCE partition function $Z_{gce}(V,T)$ is defined by the
normalization condition, $\int dE~P_{gce}(E)~=~1$. We consider a
system with fixed large volume $V$. As an illustrative example we
will use the following asymptotic form of energy distribution,
\begin{equation}\label{P_alpha_E}
P_{\alpha}(E) =
\frac{1}{\sqrt{2\pi~\omega_E~\alpha^2\overline{E}}}
 \exp \left[- ~ \frac{\left( E -\overline{E}
    \right)^2}{2\omega_E~\alpha^2~\overline{E}} \right]~,
\end{equation}
where $\overline {E}=3gVT^4/\pi^2$ is the GCE energy expectation
value and $\omega_E = (\overline{E^2} - \overline{E}^2) /
\overline{E} = 4T$ is the scaled variance of GCE energy
fluctuations. This choice for $P_{\alpha}(E)$ results in a simple
correspondence to the GCE and MCE in the large volume limit. In
Eq.~(\ref{P_alpha_E}), $\alpha$ is a dimensionless tuneable
parameter for the width of the distribution. In the MCE limit
$\alpha \rightarrow 0$, Eq.~(\ref{P_alpha_E}) becomes a Dirac
$\delta$-function, $\delta(E-\overline{E})$. For $\alpha = 1$,
Eq.~(\ref{P_alpha_E}) results in the GCE energy fluctuations
(\ref{PgceE}) in the TL \cite{clt}. The physical interpretation of
the GCE energy fluctuations corresponds to an `infinite heat
bath'. The values, $0 < \alpha < 1$, would correspond to a `large,
but finite' heat bath. The case of $\alpha > 1$ we would like to
denote as `strong' energy fluctuations.

Lastly we note that the MCE distribution $P_{mce}(N;E)$ can be conveniently
presented in terms of the GCE distributions \cite{clt}, $~ P_{mce}
(N;E) \equiv  P_{gce}(N,E) / P_{gce}(E)$. At fixed volume
$V$, the GCE distribution $P_{gce}(E)$ is given by
Eq.~(\ref{P_alpha_E}) with $\alpha =1$, while the joint GCE energy
and multiplicity distribution $P_{gce}(N,E)$ is given by a
bivariate normal distribution in the large volume limit
\cite{clt},
\eq{\label{P_E_N} & P_{gce}(N,E) \cong \frac{1}{2\pi V
\sqrt{\sigma_E^2 \sigma_N^2 \left(1-\delta^2 \right)}} \exp
\Bigg[\frac{-~1}{2V(1-\delta^2 )}
\nonumber \\
&\qquad \quad\times \left( \frac{\left( \Delta
E\right)^2}{\sigma_E^2 }
   - \frac{2 \delta ~  \Delta E \Delta N}{\sigma_E \sigma_N }
  + \frac{\left( \Delta N\right)^2}{\sigma_N^2 }
\right) \Bigg]~,}
where $\Delta E = E - \overline{E}$,~ $\Delta N = N -
\overline{N}$,~ $\overline{E} = V \kappa_1^E $,~ $\overline{N} = V
\kappa_1^N $,~ $ \sigma_E^2 = \kappa_2^{E,E} = \kappa_1^E
\omega_E$,~ $\sigma_N^2 = \kappa_2^{N,N} = \kappa_1^N \omega_{gce}
$, and the correlation coefficient $\delta = \sigma_{EN} /
(\sigma_E \sigma_N)$, with $\sigma_{EN} = \kappa_2^{E,N}$. The
relevant cumulants $\kappa$, for both $4\pi$-integrated particle
yields and for yields in limited momentum windows $\Delta p$ for
the ultra-relativistic Boltzmann gas are given in the Appendix.
One finds then for the multiplicity distribution:
\eq{
 \label{pN-alpha}  P_{\alpha} (N) &=~ \int dE~
P_{\alpha}(E)~\frac{P_{gce}(N,E)}{P_{gce}(E)}~\nonumber \\
& \cong~ \frac{1}{(2 \pi \omega_{\alpha}\overline{N})^{1/2} }
~\exp \left[ - ~\frac{\left( N-\overline {N} \right)^2}{
2 \omega_{\alpha}~\overline{N}}\right]~,  \\
 \omega_{\alpha} & =~ \omega_{mce} ~+~ \alpha^2
~\left(\omega_{gce}~-~\omega_{mce} \right)~, \label{omega-alpha}}
where  for our system $\omega_{gce}=1$ and $\omega_{mce}=1/4$. As
it can be expected, $\omega_{\alpha}=\omega_{mce}$ for $\alpha=0$,
and $\omega_{\alpha} = \omega_{gce}$ for $\alpha=1$. It also
follows, $\omega_{mce}<\omega_{\alpha}<\omega_{gce}$ for
$0<\alpha<1$, and $\omega_{\alpha}>\omega_{gce}$ for $\alpha>1$.
The $\alpha$-ensemble defined by
Eqs.~(\ref{PalphaN},\ref{P_alpha_E}) presents an extension of the
GCE ($\alpha =1$) and MCE ($\alpha =0$) to a more general energy
distribution. Different values of $\alpha$ correspond, by
construction, to the same expectation values of energy
$\overline{E}$, and multiplicity $\overline{N}$. Hence in the TL
all ensembles defined by Eq.(\ref{P_alpha_E}) are
thermodynamically equivalent. Energy and multiplicity fluctuations
are however different.


\vspace{0.2cm} As a next step we want to discuss the effect of
energy fluctuations, Eq.(\ref{P_alpha_E}), on multiplicity
fluctuations in limited segments of momentum space for a classical
ultra-relativistic gas in the TL. The results for the scaled
variances, $\omega^{\Delta p}_{\alpha}\equiv \left(\langle
N_{\Delta p}^2\rangle_{\alpha}-\langle N_{\Delta
p}\rangle^2_{\alpha}\right)/\langle N_{\Delta
p}\rangle_{\alpha}$~, in different momentum bins $\Delta p$ are
shown in Fig.~1 for several values of parameter $\alpha$. As in
Ref.\cite{acc}, each momentum bin $\Delta p = [p_1,p_2]$ contains
the same fraction $q$ of the total average multiplicity, $q\equiv
\langle N_{\Delta p}\rangle_{\alpha}/\langle N\rangle_{\alpha}$.
\begin{figure}[ht!]
\epsfig{file=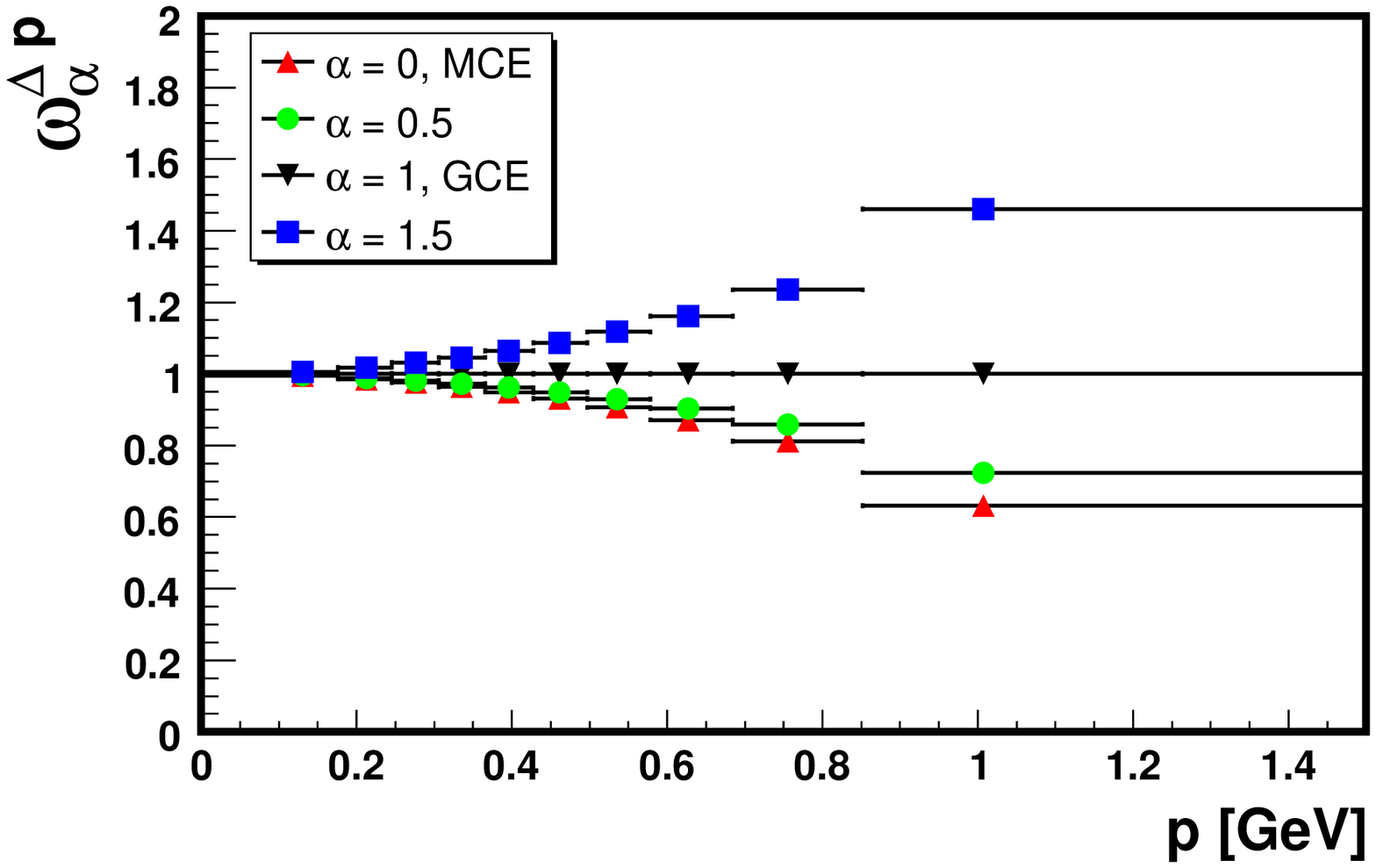,width=8cm}
\epsfig{file=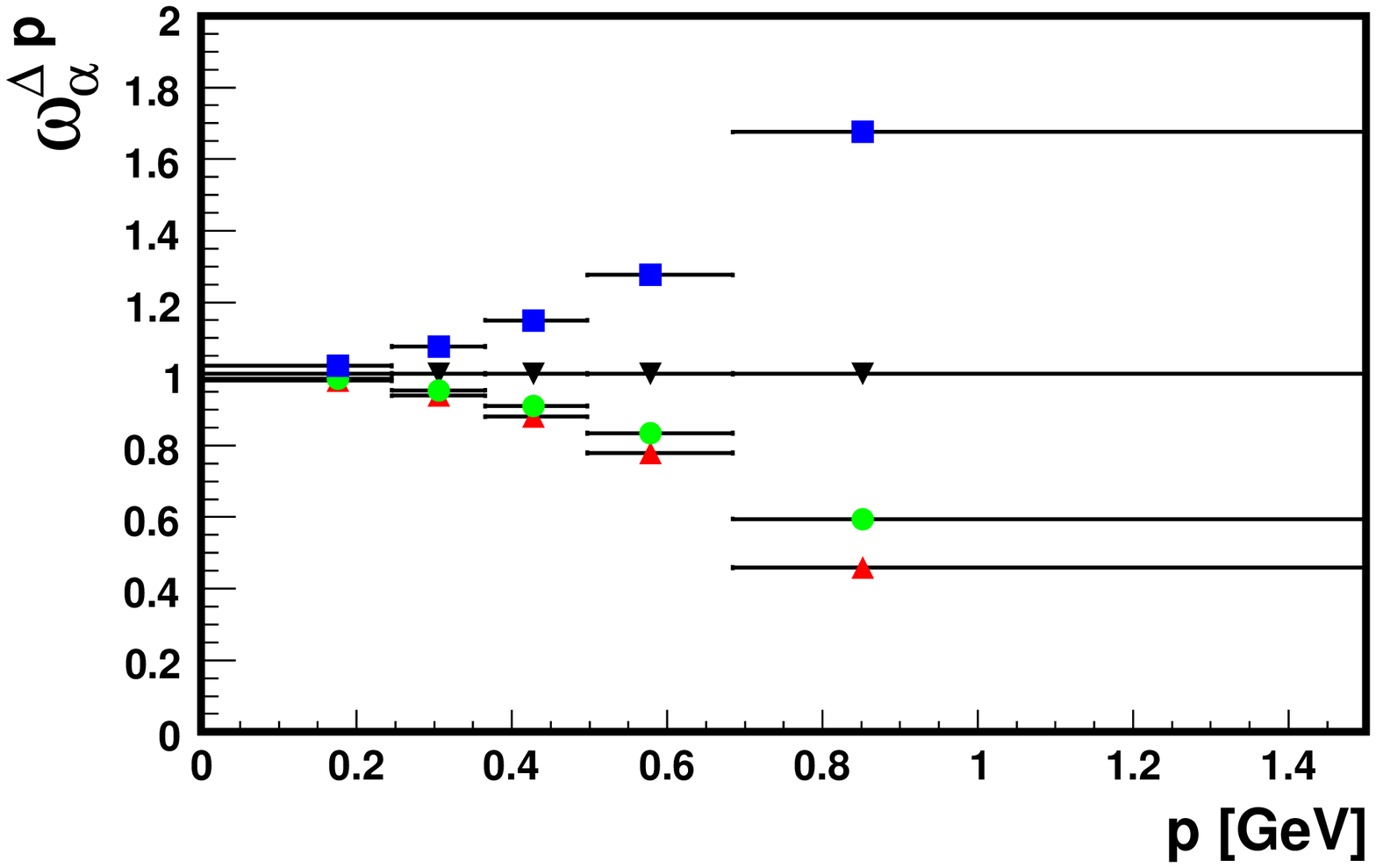,width=8cm} \caption{Momentum
dependence of the scaled variance $\omega^{\Delta p}_{\alpha}$ for
a classical ultra-relativistic gas at $T=160$~MeV. Momentum bins
are constructed in a way  that each bin contains the same fraction
$q=0.1$ (upper panel), or $q=0.2$ (lower panel) of the average
particle yield. The horizontal bars indicate the width of bins,
while the marker indicates  the center of gravity of the
corresponding bin. Calculations are done for  different values of
$\alpha$. The total particle yield would correspond to $q=1$, and
according to Eq.~(\ref{omega-alpha}) the scaled variance
$\omega_{\alpha}$ for the total yield fluctuations equals  to 1/4,
7/16, 1, 31/16 at $\alpha$=0, 0.5, 1, 1.5, respectively. }
\label{omega_alpha_plot}
\end{figure}
In the MCE ($\alpha=0$), multiplicity fluctuations are essentially
suppressed with respect to the GCE ($\alpha=1$). Global energy
conservation in the MCE introduces correlations in momentum space
~\cite{acc}. The larger the fraction of the total  energy in a
given  momentum bin, the stronger is the MCE suppression effect
($\alpha=0$ in Fig.~1). For $0<\alpha < 1$ the suppression effects
become weaker. $\alpha=1$ corresponds to the GCE results for
multiplicity fluctuations: $\omega_{\alpha}=1$, both in full
momentum space, and in different momentum bins. For `strong'
energy fluctuations, $\alpha >1$, we find an increase of
multiplicity fluctuations  with increasing bin momentum, as shown
in Fig.~1.


\vspace{0.2cm} In order to make a quantitative comparison with
present and future data on multiplicity fluctuations in
relativistic nucleus-nucleus collisions one needs to extend
essentially the formulation considered in this letter. Please note
that Eqs.(\ref{PalphaN},\ref{P_alpha_E}-\ref{P_E_N}) can be
readily generalized to more complicated cases of non-zero particle
masses, quantum statistics, many particle species, several
conserved charges, and multiplicity fluctuations in limited
segments of momentum space. In all these cases the expressions for
$P_{mce}(N;\vec{A})$ and $P_{gce}(\vec{A})$ are already obtained
\cite{clt,acc}. The MCE multiplicity distributions in systems with
full hadron-resonance spectrum and quantum statistics effects were
presented in Ref.~\cite{clt}. The role of momentum conservation
was discussed in Ref.~\cite{acc}. The next step would be an
extension of this formulation to an external distributions
$P_{\alpha}(\vec A) = P_{\alpha}(V,E, \vec{P},B,S,Q)$, e.g.,
taking into account fluctuations of energy $E$, momentum $\vec P$,
and three Abelian charges - baryon number $B$, strangeness $S$,
and electric charge $Q$. This is to be done according to
Eq.~(\ref{P}). Fluctuations and correlations of the extensive
quantities presented by the distribution $P_{\alpha}(V,E,
\vec{P},B,S,Q)$ could then be connected to measured fluctuations
of hadron multiplicities in limited segments of momentum space.
This will be a subject of future studies.


\vspace{0.2cm} In this letter we suggested to extend the concepts
of statistical ensembles and introduced the $\alpha$-ensemble
defined by an external distribution of extensive quantities. The
key assumption used for calculation of multiplicity distributions
is the `equiprobability' of all microstates with the same set of
extensive quantities. We have discussed a simple example of a gas
composed of classical massless particles to demonstrate the
effects of energy  fluctuations on  multiplicity fluctuations in
full momentum space and in limited segments of momentum space.
Measurement of event-by-event  multiplicity fluctuations in
different momentum bins correspond to current experimental studies
of hadron production in relativistic nucleus-nucleus collisions.
We believe also that the concept of statistical ensembles with
fluctuating extensive quantities may be appropriate in other
situations too. In fact, in all cases when fluctuations of
extensive quantities are a subject of interest and can be measured
experimentally.

\vspace{0.3cm} {\bf Acknowledgments.} We would like to thank V.V.~Begun,
W.~Broniowski, M.~Ga\'zdzicki, P.~Steinberg, and G.~Torrieri for fruitful
discussions. One of the authors, M.I.G., would like to thank for the support
of the Program of Fundamental Research of the Department of Physics and
Astronomy NAS of Ukraine.  

\section*{Appendix}
For the ideal Boltzmann gas of massless particles the momentum
distribution is  $f(p)=\exp(-p/T)$, and the  cumulants are:
\eq{ \kappa_1^N &= \frac{g}{2\pi^2}\int \limits_{0}^{\infty}dp~
p^2~f(p)~=~ \frac{gT^3}{\pi^2}~,\nonumber \\
 \kappa_1^E &= \frac{g}{2\pi^2}\int \limits_{0}^{\infty}dp~
p^3~f(p)~=~ 3~ \frac{gT^4}{\pi^2}~,\nonumber }
\eq{
 \kappa_2^{E,E}& = \frac{g}{2\pi^2}\int
\limits_{0}^{\infty}dp~ p^4~f(p)~=~ 12~
\frac{gT^5}{\pi^2}~\nonumber. }
 Additionally, in Boltzmann approximation, $ \kappa_2^{E,N} =
\kappa_1^{E}$ and  $\kappa_2^{N,N} =  \kappa_1^{N}$. The cumulants
in the momentum segments $\Delta p=[p_2,p_1]$ are:
\eq{
\Big( \kappa_1^N \Big)_{\Delta p}&=~ \frac{g}{2\pi^2}\int
\limits_{p_1}^{p_2}dp ~p^2~f(p) \nonumber \\
= ~ \frac{g}{2\pi^2}& \Big[~  2T^2 ~+~ 2pT ~+~ p^2 ~\Big] ~ f(p)~
 \Bigg|_{p_2}^{p_1}~,\nonumber\\
\Big( \kappa_2^{E,N} \Big)_{\Delta p} &= ~\frac{g}{2\pi^2}~\int
\limits_{p_1}^{p_2}dp ~p^3~f(p) \nonumber \\
=~ \frac{g}{2\pi^2}& \Big[ 6T^3 ~+~ 6T^2p ~ +~ 3Tp^2 ~+ ~p^3~
\Big] ~ f(p)~ \Bigg|_{p_2}^{p_1}~.\nonumber }
Additionally, in the Boltzmann approximation $( \kappa_2^{N,N})_{\Delta p} = (
\kappa_1^{N})_{\Delta p}$ .



\begin{thebibliography}{100}

\bibitem{stat1}
  J. Cleymans, H. Oeschler, K.~Redlich,
  and S.~Wheaton, Phys. Rev. C {\bf 73}, 034905 (2006);
  F.~Becattini, J.~Manninen, and~M. Ga\'zdzicki,
  {\it ibid.} {\bf 73}, 044905 (2006); A.~Andronic,
  P.~Braun-Munzinger, and J.~Stachel, Nucl. Phys. A {\bf 772}, 167
  (2006).

\bibitem{CE}
  F. Becattini, Z. Phys. C {\bf 69}, 485 (1996);
  F. Becattini and U.~Heinz, {\it ibid.} {\bf 76}, 269 (1997).

\bibitem{strange}
  J. Cleymans, K. Redlich, and E.~Suhonen, Z. Phys. C {\bf 51}, 137 (1991).

\bibitem{antibaryons}
  M.I. Gorenstein, M. Ga\'zdzicki, and W. Greiner, Phys. Lett. B
  {\bf 483}, 60 (2000).

\bibitem{charm}
  M.I. Gorenstein, A.P.~Kostyuk, H.~St\"ocker, and W.~Greiner, Phys.
  Lett. B {\bf 509}, 277 (2001).

\bibitem{MCE} F. Becattini and L. Ferroni, Eur. Phys. J. C {\bf
    35}, 243 (2004); {\bf 38}, 225 (2004); V.V.~Begun {\it et al.} J.
  Phys. G {\bf 32}, 1003 (2006).

\bibitem{fluc} V.V. Begun, M. Ga\'zdzicki, M.I.~Gorenstein, and O.S.~Zozulya,
  Phys. Rev. C {\bf 70}, 034901 (2004).

\bibitem{mce}
  V.V.~Begun, M.I.~Gorenstein, A.P.~Kostyuk, and O.S.~Zozulya,
  Phys.\ Rev.\  C {\bf 71}, 054904 (2005).

\bibitem{res}
  V.V. Begun, {\it et al.} Phys. Rev. C {\bf 74}, 044903 (2006);
  {\bf 76}, 024902 (2007).

\bibitem{RR}
Yu.B. Rumer and M.Sh. Ryvkin, {\it Thermodynamics, Statistical
Physics, and Kinetics}, Nauka, 1972 (in Russian).

\bibitem{clt}
  M.~Hauer, V.V.~Begun, and M.I.~Gorenstein, arXiv:0706.3290 [nucl-th].

\bibitem{acc}
  M.~Hauer, arXiv:0710.3938 [nucl-th].

\end{thebibliography}
\end{document}